%% file: emnlp2020.tex
\documentclass[11pt,a4paper]{article}
\usepackage[hyperref]{emnlp2020} 
\usepackage{times}
\usepackage{latexsym}
\usepackage{verbatim}
\usepackage{booktabs} 
\usepackage{algorithm}
\usepackage{algorithmic}
\usepackage{epsfig}
\usepackage{color}
\usepackage{amsmath}
\usepackage{graphicx,subfigure}
\usepackage{multirow}
\usepackage{makecell}
\usepackage{setspace}

\usepackage{microtype}

\aclfinalcopy 
\def\aclpaperid{1633} 

\title{PTUM: Pre-training User Model from Unlabeled \\User Behaviors via Self-supervision}

\author{Chuhan Wu$^\dagger$~~~~Fangzhao Wu$^\ddagger$~~~~Tao Qi$^\dagger$~~~~Jianxun Lian$^\ddagger$~~~~\textbf{Yongfeng Huang}$^\dagger$~~~~Xing Xie$^\ddagger$\\
    $^\dagger$Department of Electronic Engineering \& BNRist, Tsinghua University, Beijing 100084, China  \\
     $^\ddagger$Microsoft Research Asia, Beijing 100080, China\\
  \tt {\{wuchuhan15,wufangzhao,taoqi.qt\}@gmail.com}\\ \tt {yfhuang@tsinghua.edu.cn~~\{Jianxun.Lian, xingx\}@microsoft.com}
  }

\date{}

\date{}

\begin{document}
\maketitle

\begin{abstract}
User modeling is critical for many personalized web services.
Many existing methods model users based on their behaviors and the labeled data of target tasks.
However, these methods cannot exploit useful information in  unlabeled user behavior data, and their performance may be not optimal when labeled data is scarce.
Motivated by pre-trained language models which are pre-trained on large-scale unlabeled corpus to empower many downstream tasks, in this paper we propose to pre-train user models from large-scale unlabeled user behaviors data.
We propose two self-supervision tasks for user model pre-training.
The first one is masked behavior prediction, which can model the relatedness between historical behaviors.
The second one is next $K$ behavior prediction, which can model the relatedness between past and future behaviors.
The pre-trained user models are finetuned in downstream tasks to learn task-specific user representations.
Experimental results on two real-world datasets validate the effectiveness of our proposed user model pre-training method.

\end{abstract}

\input{data/introduction.tex}
\input{data/method.tex}

\input{data/experiment.tex}
\input{data/conclusion.tex}

\section*{Acknowledgments}

This work was supported by the National Key Research and Development Program of China under Grant number 2018YFC1604002, and the National Natural Science Foundation of China under Grant numbers U1836204, U1705261 and 61862002.
\bibliographystyle{acl_natbib}
\bibliography{emnlp2020}

\end{document}

%% file: data/introduction.tex
\section{Introduction}
User modeling is a critical technique for many personalized web services such as personalized news and video  recommendation~\cite{okura2017embedding,covington2016deep}.
Many existing methods model users from their  behaviors~\cite{zhou2018deep,ouyang2019representation}.
For example, 
Covington et al.~\shortcite{covington2016deep} proposed a YouTubeNet  model for video recommendation, which models users from their watched videos and search tokens.
Zhou et al.~\shortcite{zhou2018deep} proposed a deep interest network (DIN) for click-through rate (CTR) prediction, which models users from user behaviors on the e-commerce platform based on their relevance to the candidate ads.
Okura et al.~\shortcite{okura2017embedding} proposed to use a GRU network for news recommendation, which models users from their  clicked news.
However, these methods mainly rely on sufficient labeled data to train user models, and their performance may be not optimal  when training data is scarce.
In addition, they only model task-specific user information and do not exploit the universal user information encoded in user behaviors.

In recent years, pre-trained language models such as ELMo~\cite{peters2018deep}, BERT~\cite{devlin2019bert} and XLNET~\cite{yang2019xlnet} have achieved great success in many NLP tasks, such as reading comprehension and machine translation.
Many language models are pre-trained on a large unlabeled corpus via self-supervision tasks such as masked LM and next sentence prediction to model the contexts~\cite{devlin2019bert}.
These language models can learn universal language representations from large unlabeled corpus and empower many different downstream tasks when the labeled data for these tasks is insufficient~\cite{qiu2020pre}.

Motivated by pre-trained language models, in this paper we propose  \underline{\textbf{p}}re-\underline{\textbf{t}}rained \underline{\textbf{u}}ser \underline{\textbf{m}}odels (PTUM), which can learn universal user models from unlabeled user behaviors.\footnote{Codes and pre-trained user models are available at https://github.com/wuch15/PTUM.}
We propose two self-supervision tasks for user model pre-training.
The first one is masked behavior prediction, which aims to infer the randomly masked behavior of a user based on her other behaviors.
It can help the user model capture the relatedness between historical user behaviors.
The second one is next $K$ behaviors prediction, which aims to predict the $K$ future behaviors based on past ones.
It can help the user model capture the relatedness between past and future behaviors. 
The pre-trained user model is further fine-tuned in downstream tasks to learn task-specific user representations.
We conduct experiments on two real-world datasets for user demographic prediction and ads CTR prediction.
The results validate that our PTUM method can consistently boost the performance of many user models by pre-training them on unlabeled user behaviors.


%% file: data/method.tex
\section{Pre-trained User Model}\label{sec:Model}

\subsection{Framework of User Model}
Before introducing our PTUM method for user model pre-training,  we first briefly introduce the general framework of many existing user modeling methods based on user behaviors.
As shown in Fig.~\ref{fig.model}, the core is a behavior encoder to encode each behavior and its position into a behavior embedding and   a user encoder to learn user embeddings from behavior embeddings.
The behavior encoders can be implemented by various models.
For example, Covingon et al.~\shortcite{covington2016deep} used ID embeddings to encode watched videos and search tokens.
An et al.~\shortcite{an2019neural} used CNN to encode search queries and browsed webpages.
Wu et al.~\shortcite{wu2019nrms} used multi-head self-attention networks to encode clicked news~\cite{wu2019nrms}.
There are also many options for the user encoder, such as GRU~\cite{hidasi2016session}, attention network~\cite{wu2019} and Transformer~\cite{sun2019bert4rec}.
In these existing methods, their user models are trained in an end-to-end way using the labeled data of target task, which can only capture task-specific information. 
Thus, in this paper we propose to pre-train  user models from unlabeled user behavior data via self-supervision, which can exploit universal user information encoded in user behaviors.

\subsection{Pre-training}

We propose two self-supervision tasks for  pre-training user models on unlabeled user behaviors.
The first one is masked behavior prediction (MBP), and the second one is next $K$ behaviors prediction (NBP). 
Their details are introduced as follows:

\begin{figure}[!t]
  \centering
    \includegraphics[width=0.98\linewidth]{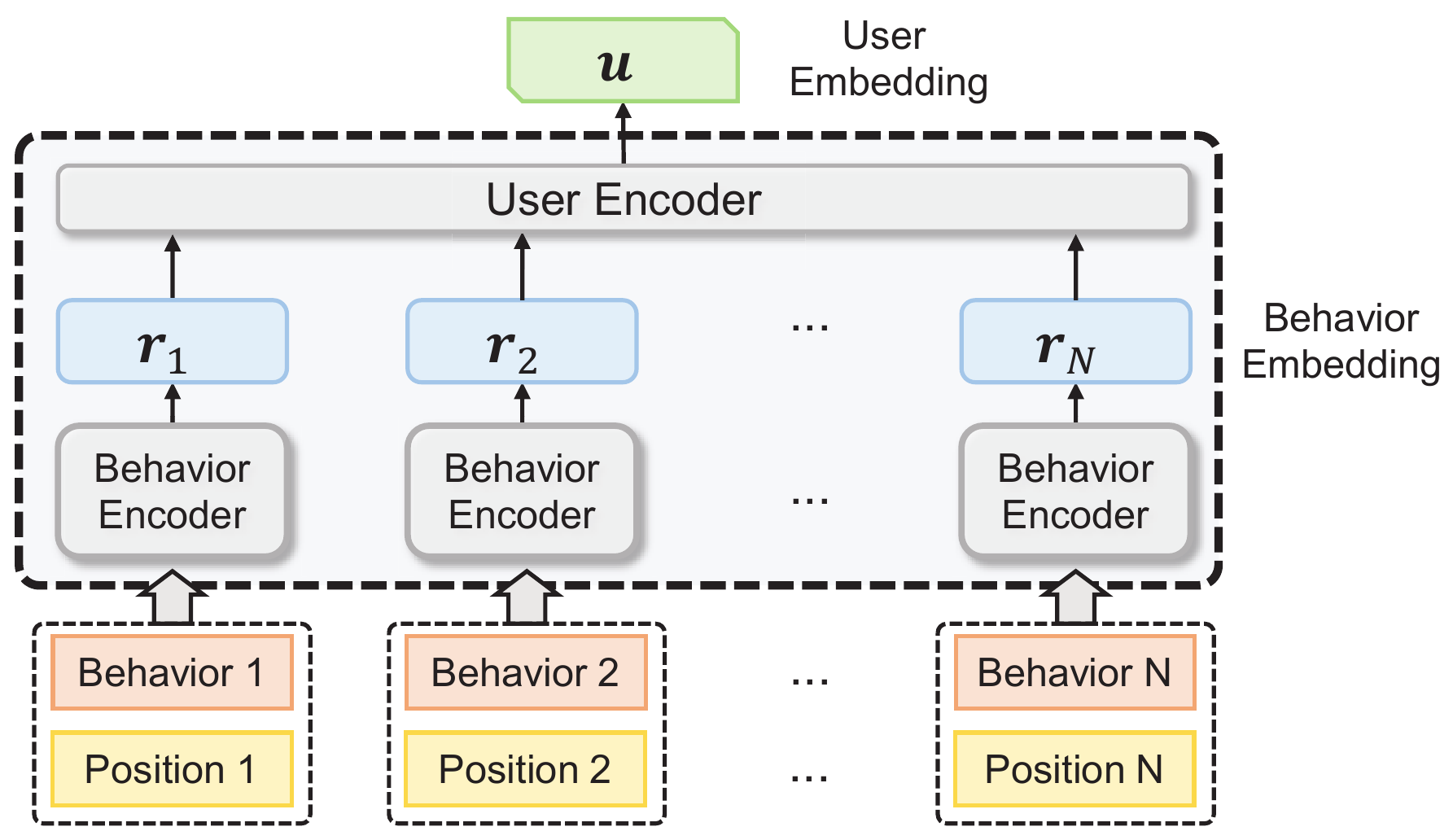}

  \caption{A general user model framework.}

  \label{fig.model}
\end{figure}

\begin{figure*}[!t]
  \centering
  \subfigure[Masked Behavior Prediction (MBP) task.]{
    \includegraphics[height=1.6in]{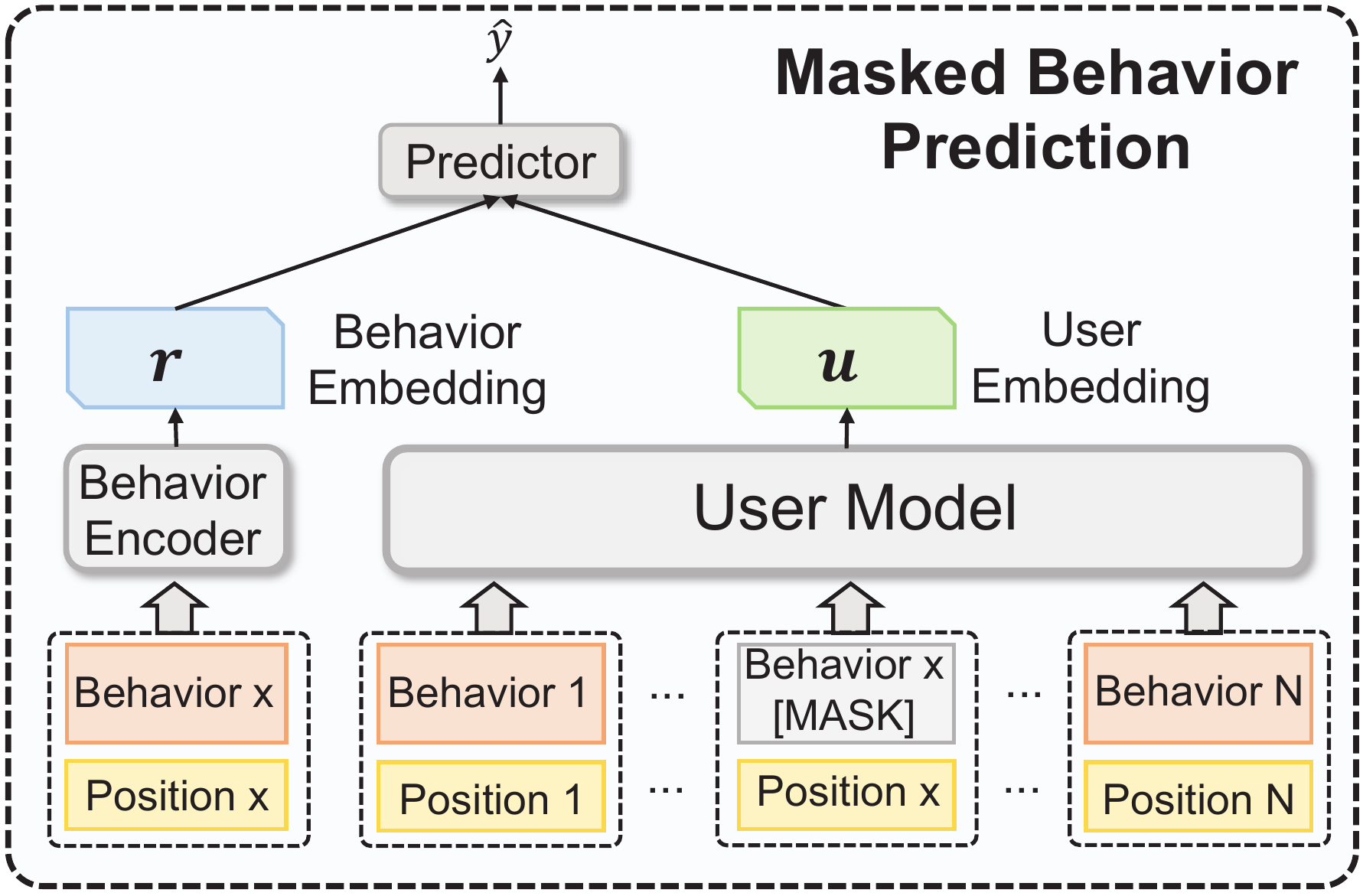}\label{task1}
    }\hspace{0.5in}
      \subfigure[Next $K$ Behaviors Prediction (NBP) task.]{
    \includegraphics[height=1.6in]{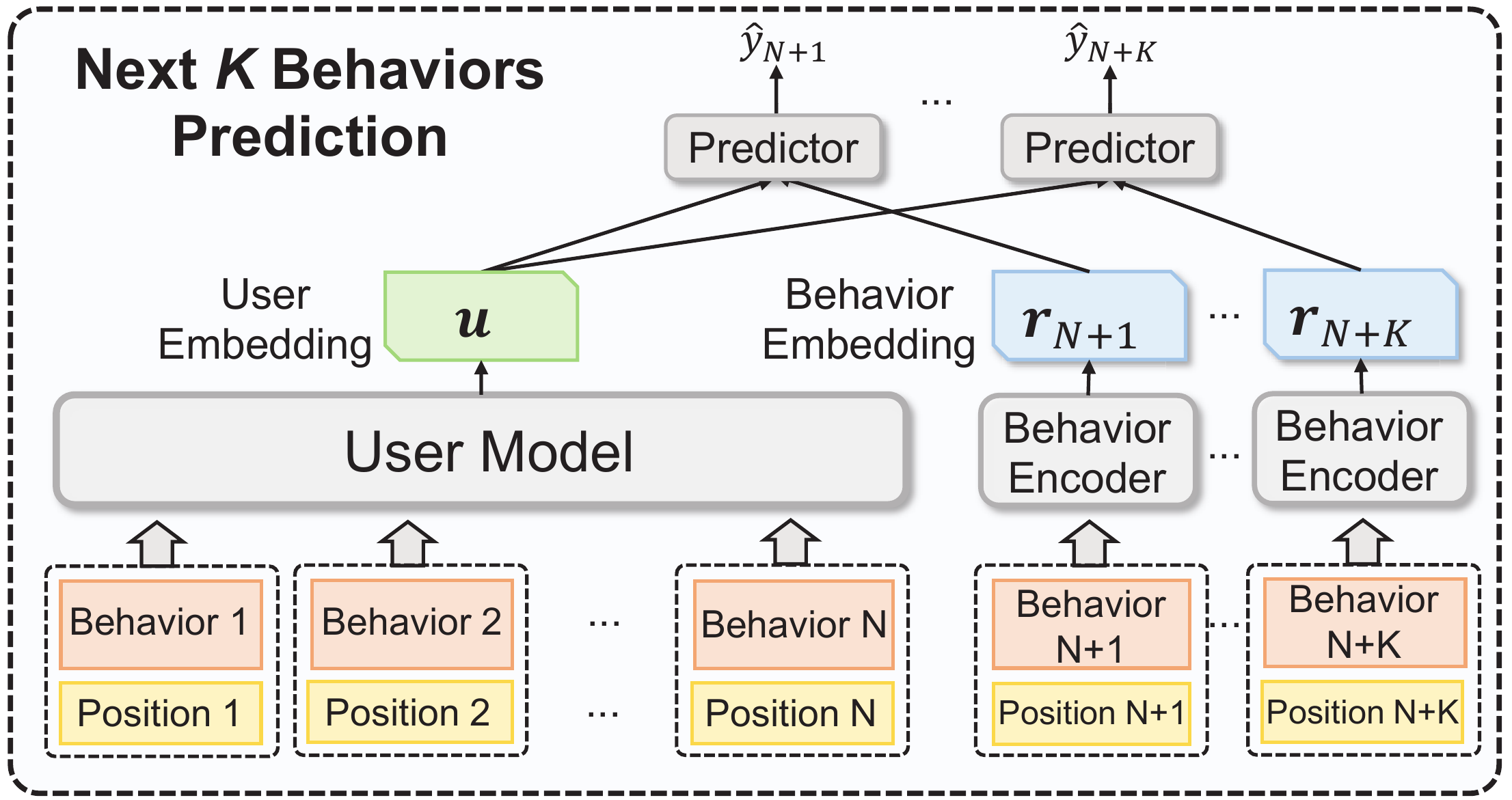}\label{task2}
    }

  \caption{Frameworks of two self-supervision tasks for user model pre-training.}  
\end{figure*}

\noindent \textbf{Task 1: Masked Behavior Prediction (MBP).}  
Modeling the relatedness between user behaviors is important for  user modeling~\cite{sun2019bert4rec}.
Inspired by the masked LM task proposed in BERT~\cite{devlin2019bert} for language model pre-training, we propose a Masked Behavior Prediction (MBP) task to pre-train user models, as shown in Fig.~\ref{task1}.
Different from words which are usually easy to be inferred from their contexts, user behaviors are diverse and are more difficult to be predicted.
Thus, different from BERT which masks a fraction of words, we only randomly mask one behavior of a user.
The goal of this task is to infer whether a candidate behavior $r$ is the masked behavior of the target user $u$ based on her other behaviors.
We use a user model to encode the behavior sequence of the user $u$ into her embedding $\mathbf{u}$, and use a behavior encoder to obtain candidate behavior embedding $\mathbf{r}$.
The relevance score $\hat{y}$ between the user $u$ and candidate behavior $r$ is evaluated by a predictor with the function $\hat{y}=f(\mathbf{u},\mathbf{r})$.

Motivated by DSSM~\cite{huang2013learning}, we use negative sampling techniques to construct self-labeled samples for user model pre-training by packing the masked behavior $r$ of a user $u$ with $P$ randomly sampled behaviors from other users.
Then, we predict the relevance scores between the user embedding and the embeddings of these $P+1$ candidate behaviors using the predictor, and normalize these scores via softmax function to obtain the probability of each candidate behavior belonging to this user.
We formulate the masked behavior prediction task as a multi-class classification problem and use the cross-entropy loss function for pre-training, which is formulated as follows:
\begin{equation}
  \mathcal{L}_{MBP}=-\sum_{y\in \mathcal{S}_1}\sum_{i=1}^{P+1}y_i\log(\hat{y}_i),
\end{equation}
where $y_i$ and $\hat{y}_i$ are the gold and predicted labels of the $i_{th}$ candidate, and $\mathcal{S}_1$ is the dataset for user model pretraining constructed from the masked behavior prediction task.

\noindent \textbf{Task 2: Next $K$ Behaviors Prediction (NBP).}  
The second self-supervision task for user model pre-training is Next $K$ Behaviors Prediction (NBP).
Modeling the relatedness between past and future behaviors is also important for user modeling~\cite{zhou2019deep}.
Thus, we propose a Next $K$ Behaviors Prediction task to help user models grasp the relatedness between past and multiple future behaviors, as shown in Fig.~\ref{task2}.
The goal is to infer whether a candidate behavior $r_{N+i}$ is the  next $i$-th behavior of the target user $u$ based on her past $N$ behaviors.
we use a user model to obtain the user embedding and use a behavior encoder to obtain the candidate behavior embeddings.
Similar to the MBP task, we use a predictor to predict the relevance score $\hat{y}_k$ between the user embedding $\mathbf{u}$ and each candidate behavior embedding $\mathbf{r}_{N+k}$.
We also use negative sampling techniques by packing each real future user behavior together with $P$ behaviors from other users to construct labeled samples for model pre-training.
The task is then formulated as $K$ parallel multi-way classification problems, and the loss function we used is  formulated as follows:

\begin{equation}\small
    \mathcal{L}_{NBP}=-\frac{1}{K}\sum_{y\in \mathcal{S}_2}\sum_{k=1}^K\sum_{i=1}^{P+1}y_{i,k}\log(\hat{y}_{i,k}),
\end{equation}
where $y_{i,k}$ and $\hat{y}_{i,k}$ are the gold and predicted labels of the $i_{th}$ candidate for the next $k_{th}$ behavior, and $\mathcal{S}_2$ is the dataset constructed from the NBP task.
We pre-train the user model in both MBP and NBP tasks collaboratively, and the final loss function to be optimized is formulated as follows:

\begin{equation}\small
    \mathcal{L} =\mathcal{L}_{MBP}+\lambda \mathcal{L}_{NBP},\label{loss}
\end{equation}
where $\lambda$ is a non-negative coefficient to control the relative importance of the NBP task.

%% file: data/experiment.tex
\section{Experiments}\label{sec:Experiments}

\subsection{Datasets and Experimental Settings}
We conduct experiments on two tasks.
The first task is user demographic prediction.
We construct a dataset (denoted as \textit{Demo}) by collecting the  webpages browsing behaviors of 20,000 users in one month (from 06/21/2019 to 07/20/2019) and their age and gender labels\footnote{Ages are categorized into 4 classes, i.e., $<$20, 20-40, 40-60 and $>$60. Gender labels have two categories.} from a commercial search engine.
The task is to infer ages and genders of users from the titles of their browsed webpages.
In this dataset, there are 12,769 male users and 7,231 female users.
There are 103 users under twenty, 2,895 users between twenty and forty, 7,453 users between forty and sixty, and 9,549 users over sixty. 
We use 80\% of users for training, 10\% for validation and the rest for test.
The second task is ads CTR prediction.
We used the dataset (denoted as \textit{CTR}) provided in~\cite{an2019neural}.
This dataset contains the titles and descriptions of ads, impression logs of ads, and the webpage browsing behaviors of 374,584 users in one month (from 01/01/2019 to 01/31/2019).
The task is to infer whether a user clicks a candidate ad based on the ad texts and the titles of browsed webpages.
We use the logs in the last week for test, and the rest for training and validation (9:1 split).
Since webpage browsing behaviors are used in both datasets, for model pre-training we use the titles of browsed webpages of 500,000 users in about six months (from 05/01/2019 to 10/26/2019), which is collected from the same platform as the \textit{Demo} dataset.
The detailed dataset statistics are shown in Table~\ref{dataset}.

\begin{table}[h]
\centering
\resizebox{0.48\textwidth}{!}{
\begin{tabular}{lrlr}
\Xhline{1.5pt}
\multicolumn{4}{c}{\textbf{Demo}}                                                    \\ \hline
\# users                    & 20,000  & avg. \# behaviors per user          & 224.7        \\
\#  behaviors                   & 4,494,771         &avg. \# words per webpage title            &   9.28  \\ \hline
\multicolumn{4}{c}{\textbf{CTR}}                                                     \\ \hline
\# users                    & 374,584 & avg. \# words per webpage title    & 10.23   \\
\# ads                      & 4,159   & avg. \# words per ad  title        & 11.95   \\
\# impressions              & 400,000 & avg. \# words per ad description  & 15.80   \\
\# clicked samples          & 364,281 & \# non-clicked samples      & 568,716  \\
\hline
\# users for pre-training          & 500,000 & \# behaviors for pre-training &   63,178,293      \\ \Xhline{1.5pt}
\end{tabular}}   
\caption{Detailed statistics of the datasets.}\label{dataset} 
\end{table}

\begin{table*}[!t]
	\centering
	\resizebox{0.98\textwidth}{!}{
\begin{tabular}{lcccccccccccc}
\Xhline{1.5pt}
\multirow{3}{*}{\textbf{Methods}} & \multicolumn{6}{c}{\textbf{Age Prediction}}                                              & \multicolumn{6}{c}{\textbf{Gender Prediction}}                                           \\ \cline{2-13} 
                         & \multicolumn{2}{c}{20\%} & \multicolumn{2}{c}{50\%} & \multicolumn{2}{c}{100\%} & \multicolumn{2}{c}{20\%} & \multicolumn{2}{c}{50\%} & \multicolumn{2}{c}{100\%} \\ \cline{2-13} 
                         & Acc.       & Macro-F     & Acc.       & Macro-F     & Acc.       & Macro-F      & Acc.       & Macro-F     & Acc.       & Macro-F     & Acc.       & Macro-F      \\ \hline
HAN                      & 51.67      & 28.40       & 53.26      & 29.54       & 55.10      & 30.78        & 69.70      & 66.54       & 72.37      & 68.64       & 73.81      & 70.20        \\
HAN+PTUM (no finetune)   & 52.16      & 28.80       & 53.62      & 29.73       & 55.29      & 30.95        & 70.12      & 66.99       & 72.59      & 68.86       & 73.91      & 70.32        \\
HAN+PTUM (finetune)      & 53.64      & 29.85       & 55.02      & 30.90       & 56.60      & 32.00        & 71.68      & 68.43       & 73.64      & 69.83       & 74.66      & 71.08        \\ \hline
HURA                     & 51.89      & 28.49       & 53.66      & 29.81       & 55.53      & 31.13        & 69.95      & 66.77       & 72.59      & 68.89       & 74.06      & 70.63        \\
HURA+PTUM (no finetune)  & 52.44      & 28.94       & 54.08      & 30.04       & 55.77      & 31.34        & 70.45      & 67.30       & 72.83      & 69.18       & 74.22      & 70.81        \\
HURA+PTUM (finetune)     & 53.88      & 29.98       & 55.46      & 31.23       & 57.09      & 32.40        & 71.95      & 68.71       & 73.95      & 70.15       & 74.92      & 71.55        \\ \hline
HSA                      & 52.25      & 28.89       & 54.13      & 30.20       & 56.27      & 31.71        & 70.16      & 66.99       & 72.96      & 69.14       & 74.65      & 71.23        \\
HSA+PTUM (no finetune)   & 52.89      & 29.41       & 54.63      & 30.52       & 56.58      & 31.99        & 70.71      & 67.55       & 73.27      & 69.50       & 74.88      & 71.48        \\
HSA+PTUM (finetune)      & \textbf{54.33}      & \textbf{30.46}       & \textbf{56.02}      & \textbf{31.70}       & \textbf{57.91}      & \textbf{33.06}        & \textbf{72.24}      & \textbf{68.98}       & \textbf{74.35}      & \textbf{70.47}       & \textbf{75.60}      & \textbf{72.24}        \\ \Xhline{1.5pt}
\end{tabular}
}

\caption{Results on the \textit{Demo} dataset under different ratios of training data. }\label{table.result}

\end{table*}

\begin{table}[!t]
	\centering
\resizebox{0.48\textwidth}{!}{
\begin{tabular}{lcccccc}
\Xhline{1.5pt}
\multirow{2}{*}{\textbf{Methods}} & \multicolumn{2}{c}{20\%}        & \multicolumn{2}{c}{50\%}        & \multicolumn{2}{c}{100\%}       \\ \cline{2-7} 
                                  & AUC            & AP             & AUC            & AP             & AUC            & AP             \\ \hline
GRU4Rec                & 71.45          & 73.20          & 71.78          & 73.85          & 72.20          & 74.40          \\
GRU4Rec+PTUM (no finetune)  & 71.76          & 73.66          & 71.95          & 74.15          & 72.33          & 74.77          \\
GRU4Rec+PTUM (finetune)        & 72.33          & 74.55          & 72.42          & 74.72          & 72.79          & 75.40          \\ \hline
NativeCTR                 & 71.64          & 73.47          & 71.96          & 74.03          & 72.35          & 74.56          \\
NativeCTR+PTUM (no finetune)   & 71.99          & 73.95          & 72.14          & 74.33          & 72.50          & 74.94          \\
NativeCTR+PTUM (finetune)         & 72.52          & 74.79          & 72.59          & 74.91          & 72.91          & 75.57          \\ \hline
BERT4Rec               & 71.82          & 73.97          & 72.39          & 74.89          & 72.99          & 75.45          \\
BERT4Rec+PTUM (no finetune) & 72.16          & 74.46          & 72.58          & 75.21          & 73.15          & 75.83          \\
BERT4Rec+PTUM (finetune)       & \textbf{72.74} & \textbf{75.34} & \textbf{73.03} & \textbf{75.81} & \textbf{73.59} & \textbf{76.48} \\ 

\Xhline{1.5pt}
\end{tabular}
}

\caption{Results on the \textit{CTR} dataset under different ratios of training data.}\label{table.result2}

\end{table}

In our experiments, the word embeddings we used were 300-dimensional.
The predictor function is implemented by dot product.
The number $K$ of future behaviors to be predicted was 2, and the coefficient $\lambda$ was 1.0.
In addition, the negative sampling ratio $P$ was 4.
These hyperparameters were tuned  on the validation data.
The complete hyperparameter settings and analysis are included in supplements.
To evaluate the performance of different methods, we used accuracy and macro F-score on the \textit{Demo} dataset, and used AUC and AP scores on the \textit{CTR} dataset.
Each experiment was repeated 10 times independently.

\subsection{Performance Evaluation}

In this section, we verify the effectiveness of our proposed PTUM method for user model pre-training.
We choose several state-of-the-art user models and compare their performance with their variants pre-trained by our PTUM method. 
On the \textit{Demo} dataset, the models to be compared include:
(1)  \textit{HAN}~\cite{yang2016hierarchical}, hierarchical attention network, which uses attentional LSTM to learn behavior  and user representations.
(2)  \textit{HURA}~\cite{wu2019neural}, hierarchical user representation with attention model, which uses CNN and attention networks to learn behavior and user representations.
(3)  \textit{HSA}~\cite{wu2019nrms}, using hierarchical multi-head self-attention to learn behavior and user representations.
On the \textit{CTR} dataset, the models to be compared include:
(1) GRU4Rec~\cite{hidasi2016session}, using GRU networks to  learn behavior and user representations.
(2) NativeCTR~\cite{an2019neural}, using CNN and attention networks to learn behavior representations and using behavior attention to learn user representations.
(3) BERT4Rec~\cite{sun2019bert4rec}, using Transformers to learn behavior and user representations.
The results on the two datasets under different ratios of training data are respectively shown in Tables~\ref{table.result} and \ref{table.result2}.
We find that pre-trained user models consistently outperform their variants trained in an end-to-end manner.
This is because pre-trained user models can capture the universal user information encoded in unlabeled user behaviors to help learn better user representations.
In addition, the advantage of pre-trained user models is larger when training data is more scarce.
This may be because pre-trained user models can exploit the complementary information provided by large-scale unlabeled user behavior data to reduce the dependency on labeled training data.
Besides, fine-tuning pre-trained user models is necessary. 
This may be because fine-tuning pre-trained user models with task-specific labeled data can help learn user representations specialized for downstream tasks.

\begin{figure}[!t]
	\centering

\subfigure[\textit{Demo} Dataset.]{\label{fig.taska}
	\includegraphics[width=0.34\textwidth]{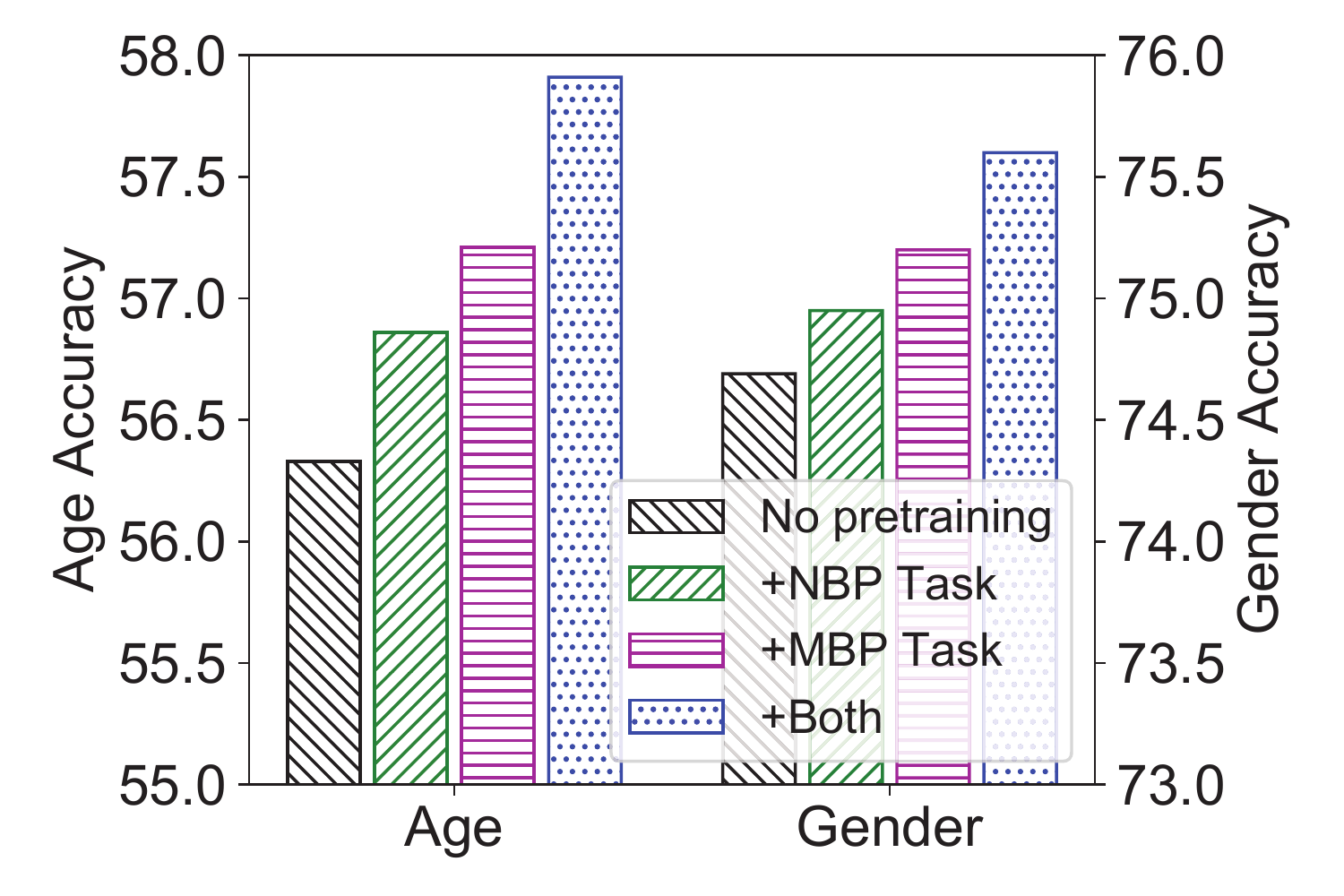}
	}
	\subfigure[\textit{CTR} Dataset.]{\label{fig.taskb}
	\includegraphics[width=0.34\textwidth]{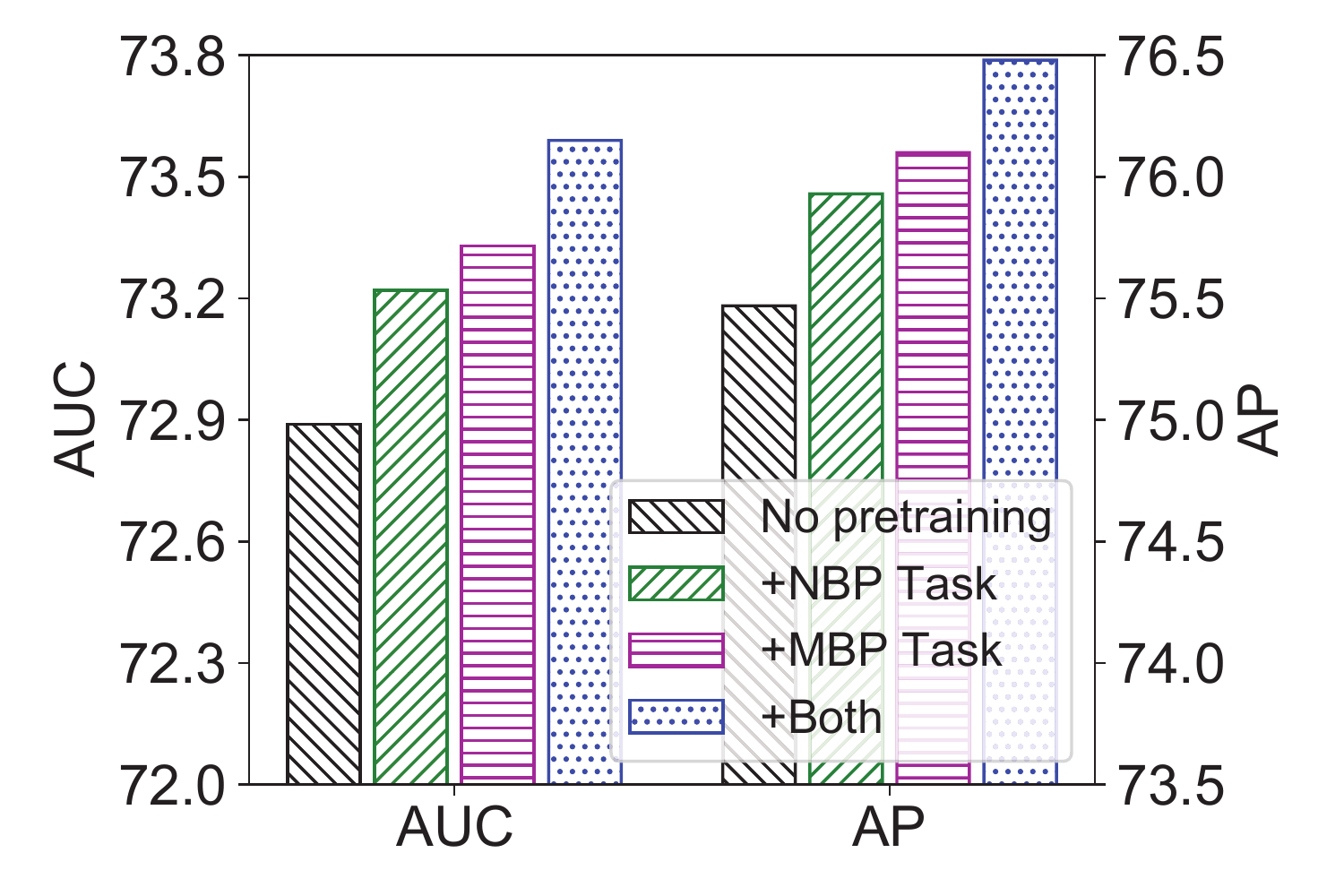}
	}  

\caption{Effect of different pre-training tasks.}\label{fig.task}  
\end{figure}

\subsection{Ablation Study}
We conducted several ablation studies to verify the effectiveness of  the proposed two self-supervision tasks for user model pre-training, i.e., masked behavior prediction and next $K$ behaviors prediction, by removing one or two of them from PTUM.
The results of \textit{HSA} on the \textit{Demo} dataset and  \textit{BERT4Rec} on the \textit{CTR} dataset are respectively shown in Figs.~\ref{fig.taska} and~\ref{fig.taskb}.
We find the masked behavior prediction task can effectively enhance pre-trained user models.
This may be because the MBP task helps user models capture the relatedness between historical user behaviors, which is critical for user modeling~\cite{sun2019bert4rec}.
In addition, the next $K$ behaviors prediction task can also improve the model performance.
This may be because the NBP task helps the user model grasp the relatedness between user behaviors in the past and future, which is also beneficial for user modeling~\cite{zhou2019deep}.
Besides, combining two tasks yields better model performance, because both the relatedness among historical behaviors and between past and future behaviors can be modeled. 

\subsection{Hyper-parameter Analysis}
\begin{figure}[t]
	\centering
\subfigure[\textit{Demo} Dataset.]{\label{fig.la}
	\includegraphics[width=0.22\textwidth]{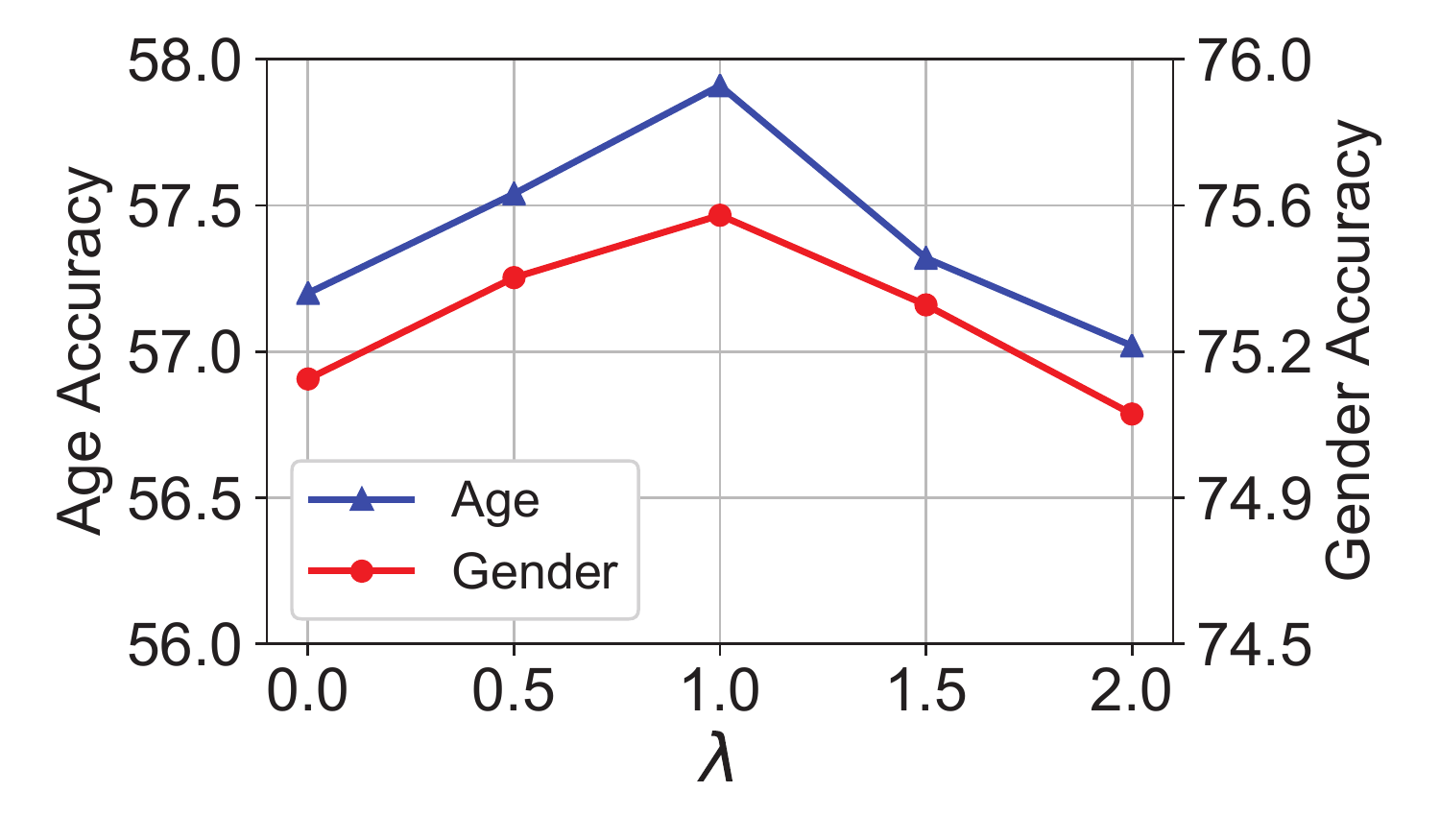}   
	}
	\subfigure[\textit{CTR} Dataset.]{\label{fig.lb}
	\includegraphics[width=0.22\textwidth]{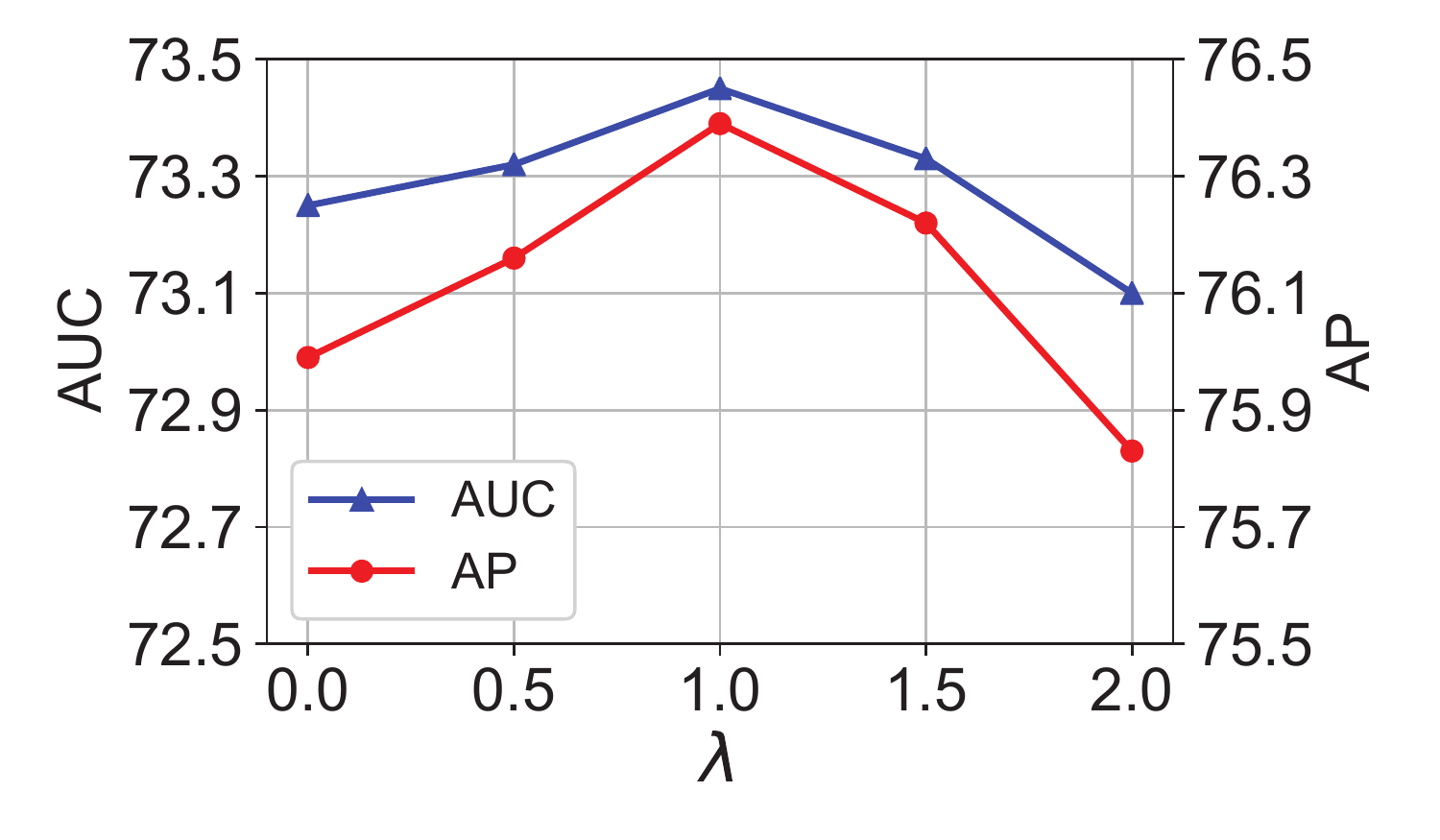}   
	}  
\caption{Performance of \textit{PTUM} w.r.t. different $\lambda$. }    
\end{figure}

In this section, we explore the influence of two key hyper-parameters on our approach, i.e., the coefficient $\lambda$ in Eq. (\ref{loss}) and the number of behavior $K$ in the NBP task.\footnote{In these experiments, the user model is \textit{HSA} on the \textit{Demo} dataset and \textit{BERT4Rec} on the \textit{CTR} dataset.}
We first vary the coefficient $\lambda$ to compare the performance of \textit{PTUM} w.r.t. different $\lambda$, and the results on the \textit{Demo} and \textit{CTR} datasets are shown in Figs.~\ref{fig.la} and~\ref{fig.lb}.
From these results, we find the performance is not optimal under a small $\lambda$.
This may be because the useful self-supervision signals in the NBP task is not fully exploited.
When $\lambda$ goes too large, the performance begins to decline.
This may be because the NBP task is over-emphasized and the MBP task is not well pre-trained.
Thus, it may be more suitable to set $\lambda=1$ to balance the two tasks.

\begin{figure}[t]
	\centering
\subfigure[\textit{Demo} Dataset.]{\label{fig.hypera}
	\includegraphics[width=0.22\textwidth]{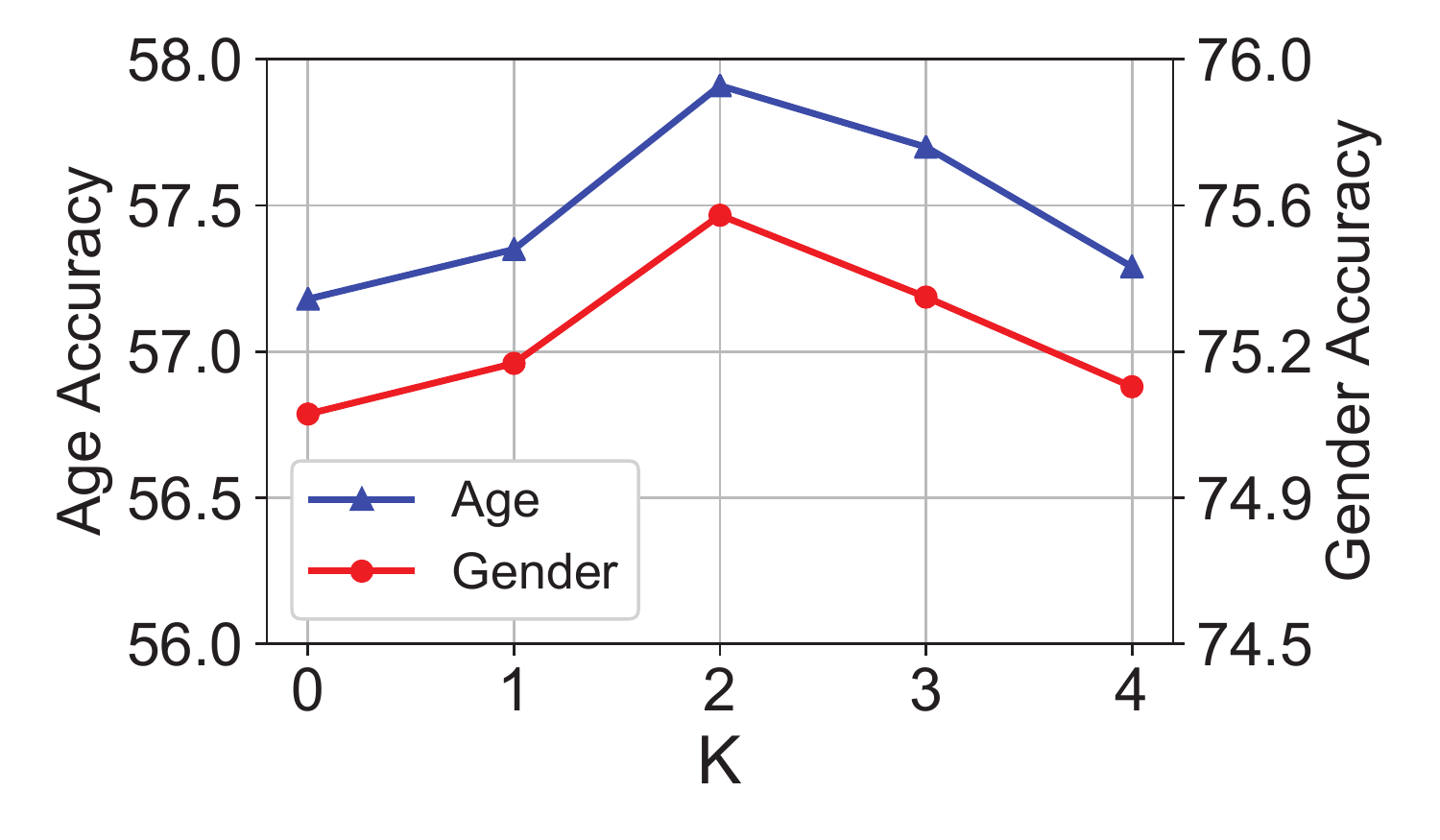}  
	}
	\subfigure[\textit{CTR} Dataset.]{\label{fig.hyperb}
	\includegraphics[width=0.22\textwidth]{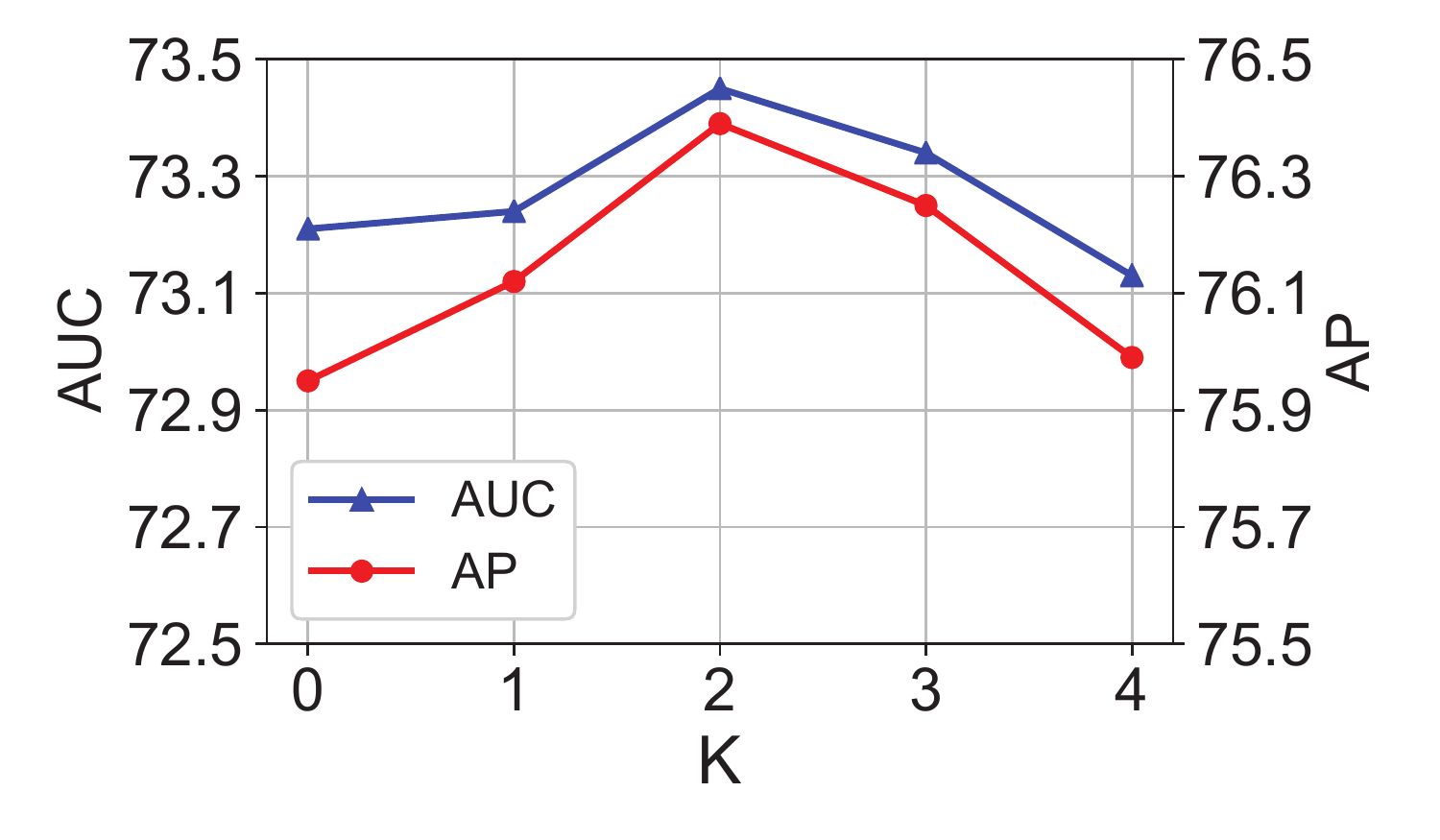}
	}  
\caption{Performance of \textit{PTUM} w.r.t. different $K$. }    
\end{figure}

Then, we vary the behavior number $K$ to explore its influence on the performance of \textit{PTUM}, and the results are shown in Figs.~\ref{fig.hypera} and~\ref{fig.hyperb}.
According to these results, we find that the performance of pre-trained user models in downstream tasks is not optimal at $K=1$.
This is probably because the relatedness between the last input behavior and the first behavior in the future may be strong, and the model may tend to overfit their short-term relatedness.
Thus, it is not optimal to simply predict the next one behavior.
In addition, we find the performance is sub-optimal when $K$ is too large.
This may be because it is difficult to accurately predict user behaviors in a long term due to the diversity of user behaviors.
Thus, a moderate $K$ may be more appropriate (e.g., $K=2$).


%% file: data/conclusion.tex
\section{Conclusion}\label{sec:Conclusion}

In this paper, we propose an effective user model pretraining method PTUM which can pretrain user models from unlabeled user behaviors.
In our method, we propose two self-supervision tasks for user model pre-training.
The first one is masked behavior prediction and the second one is next $K$ behaviors prediction, which can help user models capture the relatedness among historical behaviors and the  relatedness between past and future behaviors.
Extensive experiments on two real-world datasets for different tasks show that pre-training user models can consistently boost the performance of various user modeling methods.
